\documentclass[reprint,aps,prb]{revtex4-1}
\usepackage{natbib}
\usepackage{amsmath}
\usepackage{mathrsfs}
\usepackage{graphicx}

\begin{document}
\title{Definition of Polaritonic Fluctuations in Natural Hyperbolic Media: Applications to Hexagonal Boron Nitride, Bismuth Selenide and the Spontaneous Emission Sum Rule}
\author{Sean Molesky}
\affiliation{Department of Electrical Engineering, Princeton University, Princeton, NJ 08544, USA}
\email{smolesky@princeton.edu}
\author{Zubin Jacob}
\affiliation{School of Electrical and Computer Engineering, Brick Nanotechnology Center, Purdue University, West Lafayette, IN 47907, USA}
\affiliation{Department of Electrical and Computer Engineering, University of Alberta, Edmonton, AB T6G 2V4, CAN}
\begin{abstract}
\noindent
The discovery of photonic hyperbolic dispersion surfaces in certain van der Waals bonded solids, such as hexagonal boron nitride and bismuth selenide (a topological insulator), offers intriguing possibilities for creating strongly modified light-matter interactions. However, open problems exist in quantifying electromagnetic field fluctuations in these media, complicating typical approaches for modeling photonic characteristics. Here, we address this issue by linking the identifying traits of hyperbolic response to a coupling between longitudinal and transverse fields that can not occur in isotropic media. This description allows us to calculate the influence of hyperbolic response on electromagnetic fluctuations without explicitly imposing a characteristic size (model of nonlocality), leading to formally bounded expressions so long as material absorption is included. We then apply this framework to two exemplary areas: the optical sum rule for modified spontaneous emission enhancement in a general uniaxial medium, and thermal electromagnetic field fluctuations in hexagonal boron nitride and bismuth selenide. We find that while the sum rule is satisfied, it does not constrain the enhancement of light-matter interactions in either case. We also show that both hexagonal boron nitride and bismuth selenide possess broad spectral regions where the magnitude of electromagnetic field fluctuations are over 120 times larger, and over 800 times larger along specific angular directions, than they are in vacuum. 
\end{abstract}
\maketitle
\noindent
In 1987, Yablonovitch conceived the photonic crystal\cite{Yablonovitch1987,john1987strong} as a means of physically forbidding electromagnetic field fluctuations over a finite bandwidth to improve the performance of semiconductor lasers and solar cells.
Variations of this idea, tracing back as far as Purcell's pioneering work on nuclear magnetic resonance\cite{purcell1946modification}, appear ubiquitously in contemporary optics.
Manipulation of a system's field fluctuations characteristics (the photonic density of states, two-point correlations, ect.) provides a means of controlling an extensive list of phenomena, including field enhancement\cite{li2017balancing,khurgin2017landau}, sub-wavelength confinement\cite{arcari2014near,men2014robust,goban2015superradiance}, thermal properties\cite{Guo2012,yan2017slow,weiliang2017OE}, spontaneous\cite{hoang2015ultrafast,pelton2015modified,pick2017general} and coherent emission\cite{nowack2007coherent,wu2015monolayer}, and atom-atom interactions\cite{dong2017valley,cortes2017super}.
This strong relation between fluctuational properties and system response has naturally led to reciprocal forms of the problem originally proposed by Yablonovitch\cite{Munday2012,Belacel2013,Raman2013,Sauvan2013,lin2016enhanced}.
Namely, to what extent can electromagnetic field fluctuations inside a system be enhanced over a given spectral bandwidth?
Remarkably, realizable complements of the photonic bandgap, media in principal offering unbounded enhancement, do exist. 
Hyperbolic (indefinite\cite{smith2003electromagnetic}) media are widely stated to posses a broadband photonic dispersion singularity\cite{jacob2010engineering,jacob2012broadband} leading to electromagnetic field fluctuations of magnitude bounded only by second order effects.\\ \\
Yet, while this picture of singular (indefinite) field fluctuations has been highly successful for interpreting functional applications of hyperbolic media spanning the domains of imaging\cite{jacob2006optical,liu2007far,guo2012applications}, nanophotonics\cite{yao2011three,Ginzburg2013,Milton13,galfsky2015active,Kim2015,iorsh2015compton}, and certain semi-classical\cite{jacob2011plasmonics,newman2013enhanced,shekhar2014strong,Slobozhanyuk2015} and thermal interactions\cite{molesky2013high,jacob2014nanophotonics,Gu2014,li2015hyperbolic,dyachenko2016controlling,dai2017ultrabroadband}, there are a number of situations of interest where it can not be easily applied.
In particular, an unbounded magnitude of electromagnetic fluctuations presents difficulties for connecting semi-classical and quantum optical processes with existing results in macroscopic quantum electrodynamics\cite{milonni1995field,Barnett1996,tip1997canonical} when the fields inside a hyperbolic media must be described directly.
(As opposed to situations where boundary conditions can be used to reformulate all quantities of interest in terms of external fields.)
With the growing realization that many natural materials intrinsically exhibit high quality hyperbolic response\cite{Esslinger2014,Dai2014,Caldwell2014,Yoxall2015,Wu2015,Korzeb2015,nemilentsau2016anisotropic}, there are at least two motivations for reexamining this issue.
First, there is interest in exploring whether hyperbolic response could be useful in proposed technologies relying on quantum optical effects\cite{Cortes2012,jacob2012quantum,kapitanova2014photonic,ferrari2015hyperbolic}, which would benefit from a simplified description.
Second, as all hyperbolic media transition smoothly between frequency windows of hyperbolic and (normal) elliptic or isotropic response, the lack of a single, transparently consistent, formulation to simultaneously treat both settings is bothersome.\\ \\
\noindent
Significant steps have been made towards this goal since the start of the decade.
In 2011, Maslovski and Silveirinha\cite{maslovski2011mimicking} (wire metamaterial) and Poddubny et al.\cite{poddubny2011spontaneous} (general hyperbolic media) showed that the observable effects of the fluctuation singularity are smoothed once the emitter (atom) is given a finite size.
Like the more familiar longitudinal fluctuation divergences encountered in isotropic media, it was found that considering field interactions with a spatially finite charge distribution leads to bounds proportional to $1/a^{3}$ and $1/a$ specific to the hyperbolic regime, with $a$ denoting the characteristic size.
(For later reference, it is worthwhile to note that the particular form of the distribution was observed to have a weak influence on these terms.) 
This was shortly followed by the work of Potemkin et al.\cite{Potemkin2012}, formally rederiving the functional structure of the Green function in real space\cite{chen1983theory,Felsen1994}, in principle setting hyperbolic and isotropic media on equal footing.
The singular terms found in the expressions of this work ($\delta$ distribution terms) corrected earlier simplified guesses at the Green function singularity for uniaxial media given by Weiglhofer\cite{weiglhofer1989delta}. 
Nearly concurrently, studies by Poddubny et al.\cite{Poddubny2012} and Yan et al.\cite{Yan2012} explored the influence of microscopic models. 
Considering a cubic lattice of hyperbolically polarizable dipoles, and the interlaced metal dielectric multilayer structures frequently used for creating hyperbolic metamaterials respectively, both investigations found that nonlocal response strongly altered results based on homogeneous approximations. 
Of particular relevance to this article, the later study showed that nonlocality in the polarization response produces another mechanism bounding fluctuations in hyperbolic media, leading to a proportionality of $1/\beta^{3}$, with $\beta$ representing the length scale of nonlocality. 
(This form matches what we find for the purely longitudinal contribution, but misses the polaritonic characteristics.
Additional discussion of these developments are given in the review by Poddubny et al.\cite{Poddubny2013}.
Note that the language used in these studies, which is conventional, denotes what we will refer to as the magnitude of fluctuations or fluctuation density, as the Purcell factor or photonic density of states.
It is our belief, at least in the context of this article, that this alternate terminology will help avoid potential confusion.)\\ \\
Nevertheless, while the above findings do in fact contain all the ingredients required to characterize a general uniaxial medium, they remain unsatisfactory in practice.
Most importantly, there is no directly calculable quantity that qualifies the singular nature of the fluctuation density in the hyperbolic case, nor any obvious connection with the widely successful intuitive understanding of these media.
This information is unequivocally contained in the real space Green function as it has been previously presented, but it is not easily accessible.
(Recovering the results we will present here from real space Green function is not simple, and requires conscientious treatment of the formally singular parts of this expression.)
The two finite-size normalization approaches face similar difficulties.
While technically accurate, fluctuations in hyperbolic media appear to become inextricably tangled with the characteristic length scale.
This is both a clear break from isotropic media, where (as we will review) the transverse fluctuation density is directly determined from the permittivity parameters, and a cumbersome impediment for performing calculations in natural hyperbolic media, as in most cases only local permittivity data is readily available.          
\begin{figure*}[ht!]
  \centering
  \includegraphics{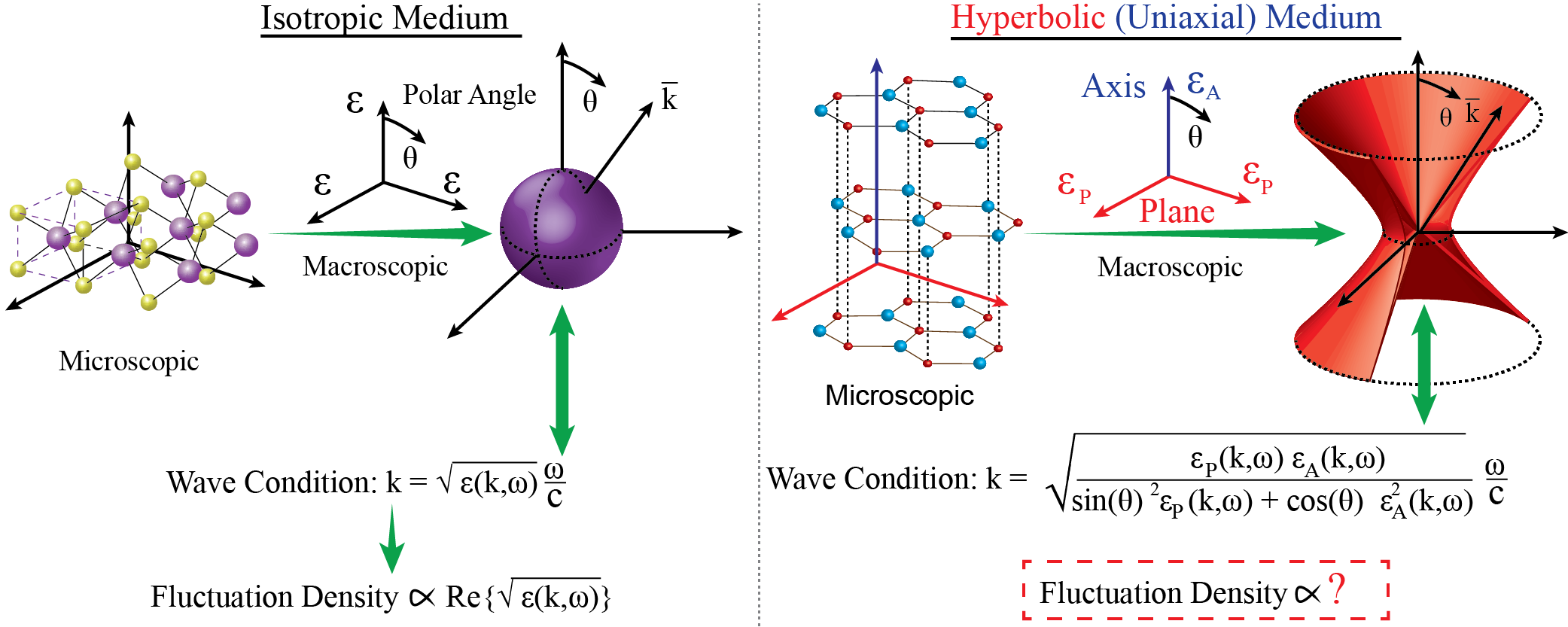}
  \vspace{0 pt}
  \caption{\textbf{Wave Conditions and Fluctuation Density}\\ In isotropic media, the magnitude of fluctuations in the electromagnetic field are proportionally related to the magnitude of the wave condition $k=\sqrt{\epsilon\left( k, \omega\right)}\;\omega / c$. By inference, the p-polarized wave condition for uniaxial media suggests that hyperbolic response should be accompanied by strongly enhanced electromagnetic field fluctuations. The central goal of this article is to quantify this statement.}
  \vspace{0 pt}
\end{figure*}
\\ \\
Here, we confront these shortcomings, consider in the plasma physics community as early as 1962\cite{arbel1962electromagnetic,fisher1969resonance}, through the essential coupling between transverse and longitudinal electric fields that arises in any anisotropic medium.
This routinely overlooked connection allows us to produce a characteristic regularized field fluctuation density bounded by material absorption, even in the hyperbolic case; and to analytically link the polariton excitations\cite{ishii2013sub} known to occur in hyperbolic media with the near-field optical and thermal properties that they exhibit.
We then apply this framework to study two areas of interest.
The first of these is the optical sum rule for modified spontaneous emission enhancement.
This general result states that the frequency integrated transverse fluctuation density is a constant of any photonic system.
(For instance, in photonic crystals the suppression of fluctuations in the band gap is compensated by a corresponding enhancement at the band edge van Hove singularities.)
Based on previously reported descriptions of the uniaxial Green function, it is not clear why this well established results continues to hold in the presence of hyperbolic response.
The second application is a calculation of the thermal fluctuation density (total electromagnetic energy density) for hexagonal boron nitride and bismuth selenide.
In these investigations, we find that while the sum rule is valid, it does not extend to the polariton modes that define hyperbolic response.
We also observe that both hexagonal boron nitride and bismuth selenide have broad spectral regions where the fluctuation density is over 120 times larger, along specific directions over 800 times larger, than it is in vacuum. 
\\ \\
\noindent
From the fluctuation dissipation theorem\cite{kubo1966fluctuation,rytov1988principles},
\begin{equation}
  \left<{\bf E}\left({\bf r},\omega\right)\otimes {\bf E}^{*}\left({\bf r}',\omega\right)\right>=\frac{\omega\;\Theta\left(\omega,T\right)}{\pi}\text{Im}\left\{\check{\mathcal{G}}\left({\bf r},{\bf r}',\omega\right)\right\},
\label{flucDiss}
\end{equation}
and our stated objective amounts to regularizing the \emph{fluctuation density} (FD) :
\begin{equation}
  \begin{split}
    \mathcal{F} \left(\omega\right) = \text{Tr}\left[\text{Im}\left\{ \check{\mathcal{G}}\left({\bf r},{\bf r}',\omega\right)\right\}\right] = \\ \int dV_{\bf k}\; \text{Tr}\left[\text{Im}\left\{e^{i\textbf{k}\cdot\left(\textbf{r}-\textbf{r}'\right)}\check{\mathcal{G}}\left({\bf k},\omega\right)\right\}\right],
  \end{split}
\label{flucDen}
\end{equation}
\noindent
for hyperbolic media.
(Where $\Theta\left(\omega , T\right)$ is the energy of a harmonic oscillator at frequency $\omega$ and temperature $T$, $\check{\mathcal{G}}\left({\bf r},{\bf r}',\omega\right)$ the dyadic Green function of the medium, $\check{\mathcal{G}}\left({\bf k},\omega\right)$ it's Fourier transform, $\int dV_{\bf k}$ an integral over reciprocal space and $\text{Im}\left\{ ... \right\}$ the imaginary part.)
The text is organized into five sections.
The first three cover our theoretical work leading to equations \eqref{GreenUni}, \eqref{GreenUniIm}, \eqref{GreenO} and \eqref{GreenE}, along with brief reviews of relevant background information.
The last two explore our chosen applications. 
\vspace{-10 pt}
\section{Polariton Excitations in Anisotropic Media}
\vspace{-10 pt}
\noindent
We begin by decomposing the Maxwell equations in terms of a chosen direction in reciprocal space vector ${\bf k}$. Letting 
\begin{align}
  {\bf w}_{_L}  & =\left(\hat{\bf k}\otimes\hat{\bf k}\right){\bf w},\tag{3a}\\
  {\bf w}_{_T} & =\left(\check{I}-\hat{\bf k}\otimes\hat{\bf k}\right) {\bf w}, \tag{3b}
\end{align}
be the projection of a vector ${\bf w}$ along $\hat{\bf k}$, and onto the plane perpendicular to $\hat{\bf k}$ respectively, any vector ${\bf w}$ can be represented as 
\begin{align}
 {\bf w}  = {\bf w}_{_L}+{\bf w}_{_T}, \tag{3c}
\end{align}
where ${\bf w}_{_L}$ and ${\bf w}_{_T}$ are referred to as the longitudinal and transverse components.
From these definitions
\begin{align}
    {\bf k}\times {\bf w}_{_L} &=0, \tag{4a}\\
  {\bf k}\cdot{\bf w}_{_T} &=0, \tag{4b}
\end{align} 
so that the Maxwell equations in vacuum separate to become
\begin{align}
  {\bf k}\cdot{\bf E}_{_L}\left({\bf k},\omega\right) & =-i\rho\left({\bf k},\omega\right)/\epsilon_{_0} \label{MaxwellF1} \tag{5a} \\
  {\bf B}_{_L}\left({\bf k},\omega\right) & =0 \label{MaxwellF2} \tag{5b} \\
  {\bf k}\times{\bf E}_{_T}\left({\bf k},\omega\right) & =\omega\;{\bf B}_{_T}\left({\bf k},\omega\right) \label{MaxwellF3} \tag{5c} \\
  ic^{2}{\bf k}\times{\bf B}_{_T}\left({\bf k},\omega\right) & =-i\omega\;{\bf E}_{_T}\left({\bf k},\omega\right)+{\bf j}_{_T}\left({\bf k},\omega\right)/\epsilon_{_0}\label{MaxwellF4} \tag{5d}\\
  i\omega\;{\bf E}_{_L}\left({\bf k},\omega\right) & ={\bf j}_{_L}\left({\bf k},\omega\right)/\epsilon_{_0}. \nonumber
\end{align}
with $\epsilon_{_0}$ and $\mu_{_0}$ denoting the permittivity and permeability of vacuum, $\rho\left({\bf k},\omega\right)$ the charge density, ${\bf j}\left({\bf k},\omega\right)$ the current density, ${\bf B}\left({\bf k},\omega\right)$ the magnetic field, and ${\bf E}\left({\bf k},\omega\right)$ the electric field.
Assuming that the relative permeability is negligibly different than vacuum, $\check{\mu}\left({\bf k},\omega\right)=\check{I}$, as we will throughout, macroscopic averaging of (5a)-(5d) produces
\begin{align}
  {\bf k}\cdot{\bf D}_{_L}\left({\bf k},\omega\right) & =-i\rho_{f}\left({\bf k},\omega\right)/\epsilon_{_0} \tag{6a} \label{M1}\\
  \bar{\bf B}_{_L}\left({\bf k},\omega\right) & = 0 \tag{6b} \label{M2} \\
  {\bf k}\times\bar{\bf E}_{_T}\left({\bf k},\omega\right) & =\omega\;\bar{\bf B}_{_T}\left({\bf k},\omega\right) \tag{6c} \label{M3} \\
  ic^{2}{\bf k}\times\bar{\bf B}_{_T}\left({\bf k},\omega\right) & =-i\omega\;{\bf D}_{_T}\left({\bf k},\omega\right)+\;{\bf j}_{f_T}\left({\bf k},\omega\right)/\epsilon_{_0} \label{M4} \tag{6d}\\
  i\omega\;{\bf D}_{_L}\left({\bf k},\omega\right) & ={\bf j}_{f_L}\left({\bf k},\omega\right)/\epsilon_{_0} \nonumber. 
\end{align}
where ${\bf D}\left({\bf k},\omega\right)=\check{\epsilon}\left(k,\omega\right){\bf E}\left({\bf k},\omega\right)$ is the electric displacement field, and the $f$ subscript is introduced as a shorthand to mark that the quantity is free,
i.e. separate from the microscopic densities that have been averaged over in producing (6a)-(6d) from (5a)-(5d). 
The overline $\bar{X}$ serves as a similar reminder that the electric and magnetic fields appearing in (6a)-(6d) are spatially averaged, and not equivalent to the identically named fields in (5a)-(5d). 
\vspace{-10 pt}
\subsection{Microscopic densities} 
\vspace{-10 pt} 
\noindent
The microscopic equations, (5a) and (5b), show that the longitudinal electric and magnetic fields are entirely determined by their respective charge densities,
\begin{align}
 {\bf E}_{_L}\left({\bf k},\omega\right) & =-i\rho\left({\bf k},\omega\right){\bf k}/\left(\epsilon_{_0}k^{2}\right) \tag{7a} \\
 {\bf B}_{_L}\left({\bf k},\omega\right) & =0,\tag{7b}
\end{align}
so that they vanish in the absence of charge. (Since the most important properties related to these quantities are the Coulomb self-energy and charge momentum\cite{cohen1997photons}, we interchangeably refer to them as \emph{Coulombic}.) A homogeneous solution to (6a)-(6d) is hence purely transverse, and from (6c), determined by ${\bf E}_{_T}\left({\bf k},\omega\right)$. (We refer to all such quantities that define homogeneous solutions as \emph{normal variables}.)
\vspace{-10 pt}
\subsection{Isotropic media} 
\vspace{-10 pt} 
\noindent
A treatment of isotropic media follows largely by analogy. 
From the relation between ${\bf D}\left({\bf k},\omega\right)$ and $\bar{\bf E}\left({\bf k},\omega\right)$, defined by the scalar relative permittivity $\check{\epsilon}\left( k,\omega\right)=\epsilon\left( k,\omega\right)I_{3}$, the longitudinal electric and magnetic fields are
\begin{align}
  \bar{\bf E}_{_L}\left({\bf k},\omega\right) & =-i\rho_{f}\left({\bf k},\omega\right){\bf k}/\left(k^{2}\epsilon_{_0}\epsilon\left( k,\omega\right)\right), \tag{8a}\\
    \bar{\bf B}_{_L}\left({\bf k},\omega\right) & =0. \tag{8b}
\end{align}
Therefore, so long as $\epsilon\left( k,\omega\right)\neq 0$, the normal variables of an isotropic medium are again transverse and equivalent to $\bar{\bf E}_{_T}\left({\bf k},\omega\right)$. \\ \\
\noindent
The caveat to this congruence is the appearance of the polarization condition 
\setcounter{equation}{8}
\begin{equation}
  \epsilon\left(k,\omega\right)=0
  \label{charge}
\end{equation} in equation (6a). 
When this condition is met, the displacement field may be zero even if the longitudinal electric field is not. 
As the remaining macroscopic equations do not depend on the longitudinal electric field, these Coulombic modes evolve independent of the transverse electromagnetic solutions\cite{Ferrell1958,Berreman1963,Newman2015}. 
Averaging (5a) directly  
\begin{equation}
   \bar{\rho}\left({\bf k},\omega\right)=i\epsilon_{_0} k\;\cdot\bar{E}_{_L}\left({\bf k},\omega\right), 
\end{equation}
showing that each Coulombic solution is a mechanical macroscopic oscillation of the microscopic charge density, mediated by the electric field. \\ \\
\noindent 
The appearance of these Coulombic solutions make (6a)-(6d) fundamentally different than a scaled vacuum.
The fact that $\epsilon\left({\bf k}, \omega\right)$ exists because of the presence of charges is inescapable, even after macroscopic averaging.
However, the effects resulting from the two solution types can usually be treated independently as they tend to exhibit markedly different behaviour.
\vspace{-10 pt}
\subsection{Anisotropic (uniaxial) media} 
\vspace{-10 pt} 
\noindent
For anisotropic media the relative permittivity tensor, $\check{\epsilon}\left( k,\omega\right)$, can not simplified and a more careful analysis is required. 
Rewriting (6a)-(6d) in the Coulomb gauge,
\begin{align}
    \bar{\bf E}_{_T}\left({\bf k},\omega\right) & =i\omega\bar{{\bf A}}_{_T}\left({\bf k},\omega\right)\tag{11a}\\
    \bar{\bf E}_{_L}\left({\bf k},\omega\right) & =-i{\bf k}\;\bar{V}\left({\bf k},\omega\right)\tag{11b}\\
  \bar{\bf B}_{_T}\left({\bf k},\omega\right) & =i{\bf k}\times\bar{{\bf A}}_{_T}\left({\bf k},\omega\right)\tag{11c}
\end{align}
a homogeneous solution requires both
\setcounter{equation}{11}
\begin{equation}
  \begin{split}
    \omega\left(I_{3} k^2-{\bf k}\otimes{\bf k}-k_{_0}^2\;\check{\epsilon}\left( k,\omega\right)\right)\bar{{\bf A}}_{_T}\left({\bf k},\omega\right)+\\
    k^{2}_{_0}\check{\epsilon}\left(k,\omega\right){\bf k}\;\bar{V}\left({\bf k},\omega\right)=0,
  \end{split}
  \label{C1}
\end{equation}
and 
\begin{equation}
  {\bf k}\;\check{\epsilon}\left( k,\omega\right)\left(\omega\;\bar{{\bf A}}_{_T}\left({\bf k},\omega\right)-{\bf k}\;\bar{V}\left({\bf k},\omega\right)\right)=0,
  \label{C2}
\end{equation}
with $k_{_0}=\omega/c$, $\bar{{\bf A}}\left({\bf k},\omega\right)$ standing for the electromagnetic vector potential, and $\bar{V}\left({\bf k},\omega\right)$ the scalar potential. 
In order to satisfy \eqref{C2} there are three distinct possibilities.\\ \\
\noindent
(O): If in addition to being perpendicular to ${\bf k}$, $\bar{{\bf A}}_{_T}\left({\bf k},\omega\right)$ is constrained to directions perpendicular to ${\bf k}\;\check{\epsilon}\left(k,\omega\right)$ then $\bar{V}\left({\bf k},\omega\right)=0$. 
As in the microscopic picture, the normal variables are then transverse. 
Evaluating \eqref{C1}, simplifying to a uniaxial response as we will throughout, produces the s-polarized, or \emph{ordinary}, wave condition 
\begin{equation}
  k=\sqrt{\epsilon_{_P}\left( k,\omega\right)}\;k_{_0},
  \label{ordCond}
\end{equation}
with $\bar{{\bf A}}_{_T}\left({\bf k},\omega\right)$ confined to the direction $\hat{\bf s}=\left[-s\left(\phi\right),c\left(\phi\right),0\right]$ relative to the unit direction in reciprocal space $\hat{\bf k}=\left[s\left(\theta\right)c\left(\phi\right),s\left(\theta\right)c\left(\phi\right),c\left(\theta\right)\right]$. 
(Our labeling convention for uniaxial media is shown in Fig.1.)\\ \\
\noindent
(C): If ${\bf k}\;\check{\epsilon}\left( k,\omega\right){\bf k}=0$ then $\bar{V}\left({\bf k},\omega\right)$ can be non-zero independent of the value of $\bar{{\bf A}}_{_T}\left({\bf k},\omega\right)$. 
These purely longitudinal solutions generalize the Coulombic modes of an isotropic media, \eqref{charge}, with the updated criterion
\begin{equation}
  {\bf k}\;\check{\epsilon}\left( k,\omega\right){\bf k}=0,
  \label{polAnsio}
\end{equation}
accounting for the reduced symmetry of $\check{\epsilon}\left( k,\omega\right)$. 
For uniaxial anisotropy, \eqref{polAnsio} reduces to 
\begin{equation}
  \epsilon_{_U}\left( k,\theta,\omega\right)=0,
  \label{cCond}
\end{equation}
with 
\begin{equation}
  \epsilon_{_U}\left( k,\theta,\omega\right)=s\left(\theta\right)^{2}\epsilon_{_P}\left( k,\omega\right)+c\left(\theta\right)^{2}\epsilon_{_A}\left( k,\omega\right).
  \label{polUni}
\end{equation}
We will refer to this projection as the \emph{uniaxial permittivity} of the medium.
\\ \\
\noindent
(AP): If $\bar{{\bf A}}_{_T}\left({\bf k},\omega\right)$ is not perpendicular to ${\bf k}\;\check{\epsilon}\left( k,\omega\right)$, then \eqref{C2} forces the equality 
\begin{equation}
  \bar{V}\left({\bf k},\omega\right)=\omega\;{\bf k}\;\check{\epsilon}\left( k,\omega\right)\;\bar{{\bf A}}_{_T}\left({\bf k},\omega\right)/\left({\bf k}\;\check{\epsilon}\left( k,\omega\right){\bf k}\right)
\end{equation}
and \eqref{C1} becomes  
\begin{align}
  &\left(\check{I}\frac{k^{2}}{k_{_0}^{2}}-\frac{{\bf k}\otimes{\bf k}}{k_{_0}^{2}}-\check{\epsilon}\left( k,\omega\right)+\frac{\check{\epsilon}\left( k,\omega\right){\bf k}\otimes\check{\epsilon}\left( k,\omega\right){\bf k}}{{\bf k}\;\check{\epsilon}\left( k,\omega\right){\bf k}}\right)\bar{{\bf A}}_{_T}= \nonumber \\
  &0.
\end{align}
Given the directional restrictions on $\bar{{\bf A}}_{_T}$, satisfaction of this constraint requires $k$ to be a solution of the p-polarized, or \emph{extraordinary}, wave condition 
\begin{equation}
  k=\sqrt{\epsilon_{_E}\left(k,\theta,\omega\right)}\;k_{_0},
  \label{eoCond}
\end{equation} 
with $\bar{{\bf A}}_{_T}\left({\bf k},\omega\right)$ confined to the direction $\hat{\bf p}=$ $\left[-c\left(\theta\right)c\left(\phi\right),-c\left(\theta\right)s\left(\phi\right),s\left(\theta\right)\right]$, and $\epsilon_{_E}\left(\bf k,\theta,\omega\right)$ defined as the \emph{extraordinary permittivity}
\begin{equation}
  \epsilon_{_E}\left(k,\theta,\omega\right)=\frac{\epsilon_{_A}\left( k,\omega\right)\epsilon_{_P}\left( k,\omega\right)}{\epsilon_{_U}\left(k,\theta,\omega\right)}.
  \label{polE}
\end{equation}
\noindent
(The similarity between \eqref{polE} and the excitation condition of a surface plasmon polariton\cite{ford1984electromagnetic} is not coincidental.)\\ \\ 
Substitution into $\bar{V}\left({\bf k},\omega\right)=\omega{\bf k}\;\check{\epsilon}\left( k,\omega\right)\bar{{\bf A}}_{_T}\left({\bf k},\omega\right)/$ $\left({\bf k}\;\check{\epsilon}\left(k,\omega\right){\bf k}\right)$ shows that for the extraordinary family of solutions
\begin{equation}
  \begin{split}
\bar{\bf E}_{_{L}}\left({\bf k},\omega\right)=\hat{{\bf k}}\;
\epsilon_{_H}\left( k,\theta,\omega\right)\bar{E}_{_T}\left({\bf k},\omega\right), 
  \end{split}
\label{ANISOELong}
\end{equation}
where $\bar{{E}}_{_T}\left({\bf k},\omega\right)$ is the undetermined scalar magnitude of the transverse component of the electric field,
\begin{equation}
  \epsilon_{_\Delta}\left( k,\theta,\omega\right)=s\left(\theta\right)c\left(\theta\right)\left(\epsilon_{_P}\left( k,\omega\right)-\epsilon_{_A}\left( k,\omega\right)\right)
  \label{polAniso}
\end{equation} 
is the relative degree of polarization anisotropy between the optical axis and plane, and 
\begin{equation}
  \epsilon_{_H}\left( k,\theta,\omega\right) = \frac{\epsilon_{_\Delta}\left( k,\theta,\omega\right)}{\epsilon_{_U}\left( k,\theta,\omega\right)}
    \label{hypPer}
\end{equation}
defines the \emph{hyperbolic permittivity}.\\ \\
\noindent
\eqref{ANISOELong} contains the essential physics that will guide the rest of the manuscript: in an anisotropic media the normal variables are a mixture of transverse and longitudinal fields. Averaging (5a) as before,
\begin{equation}
  \begin{split}
\bar{\rho}\left({\bf k},\omega\right)=i\epsilon_{_0} k\;
\epsilon_{_H}\left( k,\theta,\omega\right)\bar{E}_{_T}\left({\bf k},\omega\right),
  \end{split}
\label{ANISOqLong}
\end{equation}
making it is apparent that for any extraordinary solution the electromagnetic (transverse) oscillation is accompanied by a Coulombic charge oscillation.
In light of this fundamental coupling, we refer to these solutions as \emph{anisotropic polaritons} (AP). 
(Notice that the same analysis can be applied to the magnetic field and relative permeability tensor $\check{\mu}\left( k,\omega\right)$.)
From B., it is clear that in isotropic media such excitations are impossible. 
The global direction of the electric field for a Coulombic mode is uniquely fixed by the propagation direction of the charge oscillation.
This means that longitudinal electric fields can not couple to magnetic fields, and hence can not be electromagnetic.
\eqref{hypPer} and \eqref{ANISOqLong} show that the Coulombic part of an AP type mode grows proportionally with the degree of anisotropy of the medium, \eqref{polAniso}, and is resonant with zeros of the uniaxial permittivity, \eqref{polUni}. Accordingly, these properties also characterize the distinguishing features of hyperbolic response. 
\vspace{-10 pt}
\section{Normal Variable Decomposition of the Anisotropic Green Function}
\vspace{-10 pt}
\noindent
Substituting (6c) into (6d), the electric field inside an anisotropic medium obeys the equation 
\begin{align}
&-{\bf k}\times{\bf k}\times\bar{\bf E}\left({\bf k},\omega\right)-k_{_0}^{2}\check{\epsilon}\left(k,\omega\right)\bar{\bf E}\left({\bf k},\omega\right)= \nonumber \\
&i {\bf j}_{f}\left({\bf k},\omega\right)/\left(\epsilon_{_0}c^{2}\right),\nonumber \\
&\epsilon_{_0}c^{2}\left(
k^2\left(I_{3}-\hat{\bf k}\otimes\hat{\bf k}\right)-k_{_0}^{2}\;\check{\epsilon}\left( k,\omega\right)\right)\bar{\bf E}\left({\bf k},\omega\right)= \nonumber\\ 
&\check{\mathcal{G}}^{-1}\left({\bf k},\omega\right)\bar{\bf E}\left({\bf k},\omega\right)=i {\bf j}_{f}\left({\bf k},\omega\right).
\label{GreenInverse}
\end{align}
The dyadic Green function of a uniaxial medium is the formal inverse of this relation,
\begin{align}
&\bar{\bf E}\left({\bf k},\omega\right)=i \check{\mathcal{G}}\left({\bf k},\omega\right) {\bf j}_{f}\left({\bf k},\omega\right),\nonumber \\
&\check{\mathcal{G}}^{_U}\left({\bf k},\omega\right)=\frac{k_{_0}}{\epsilon_{_0}c^{2}}\Bigg(\frac{\hat{{\bf s}}\otimes\hat{{\bf s}}}{k^2-\epsilon_{_P}\left( k,\omega\right)}-\frac{\hat{\bf k}\otimes\hat{\bf k}}{\epsilon_{_U}\left( k,\theta,\omega\right)}+\nonumber\\
&\frac{\left(\hat{\bf p}+\epsilon_{_H}\left( k,\theta,\omega\right)\;\hat{\bf k}\right)\otimes\left(\hat{\bf p}+\epsilon_{_H}\;\left( k,\theta,\omega\right)\hat{\bf k}\right)}{k^2-\epsilon_{_E}\left( k,\theta,\omega\right)}\Bigg),
\label{GreenUni}
\end{align}
where all reciprocal vectors have been normalized by $k_{_0}$, and, recalling our previous definitions,
\begin{align}
  &\hat{{\bf s}} =\left[-s\left(\phi\right),c\left(\phi\right),0\right]\tag{28a}\\
  &\hat{{\bf p}} =\left[-c\left(\theta\right)c\left(\phi\right),-c\left(\theta\right)s\left(\phi\right),s\left(\theta\right)\right]\tag{28b}\\
  &\hat{{\bf k}} =\left[s\left(\theta\right)c\left(\phi\right),s\left(\theta\right)s\left(\phi\right),c\left(\theta\right)\right]\tag{28c}
\end{align}
\begin{align}
&\epsilon_{_U}\left( k,\theta,\omega\right)=s\left(\theta\right)^{2}\epsilon_{_P}\left( k,\omega\right)+c\left(\theta\right)^{2}\epsilon_{_A}\left( k,\omega\right)\tag{29a}\\
& \epsilon_{_E}\left(k,\theta,\omega\right)=\frac{\epsilon_{_A}\left( k,\omega\right)\epsilon_{_P}\left( k,\omega\right)}{\epsilon_{_U}\left(k,\theta,\omega\right)}
 \tag{29b}\\
&\epsilon_{_\Delta}\left( k,\theta,\omega\right)=s\left(\theta\right)c\left(\theta\right)\left(\epsilon_{_P}\left( k,\omega\right)-\epsilon_{_A}\left( k,\omega\right)\right)\tag{29c}\\
&\epsilon_{_H}\left( k,\theta,\omega\right) = \frac{\epsilon_{_\Delta}\left( k,\theta,\omega\right)}{\epsilon_{_U}\left( k,\theta,\omega\right)}\tag{29d}.
\end{align}
In isotropic media, $\epsilon_{_H}\left( k,\omega\right)$ reduces to zero while $\epsilon_{_U}\left( k,\omega\right)$ and $\epsilon_{_E}\left( k,\omega\right)$ become the isotropic permittivity $\epsilon\left( k,\omega\right)$ so that \eqref{GreenUni} simplifies to
\setcounter{equation}{29}
\begin{align}
&\check{\mathcal{G}}^{_I}\left({\bf k},\omega\right)=\nonumber\\
&\frac{k_{_0}}{\epsilon_{_0}c^{2}}\left(\frac{\hat{{\bf s}}\otimes\hat{{\bf s}}}{k^2-\epsilon\left( k,\omega\right)}-\frac{\hat{\bf k}\otimes\hat{\bf k}}{\epsilon\left(k,\omega\right)}+\frac{\hat{\bf p}\otimes\hat{\bf p}}{k^2-\epsilon\left( k,\omega\right)}\right).
  \label{GreenIso}
\end{align}
(For the remainder of the article, the $_U$ and $_I$ superscripts mark that the results applies specifically to either uniaxial or isotropic media.)\\ \\
\noindent
As an operator, the Green function \eqref{GreenUni} determines the electric field generated by a point current source as a modal expansion of the three homogeneous solution families.\\ \\ 
(O): Ordinary electromagnetic 
\begin{equation}
\check{\mathcal{G}}^{_U}_{_O}\left({\bf k},\omega\right)=\frac{k_{_0}}{\epsilon_{_0}c^{2}}\;\frac{\hat{{\bf s}}\otimes\hat{{\bf s}}}{k^2-\epsilon_{_P}\left( k,\omega\right)}.
\end{equation}
(C): Coulombic  
\begin{equation}
\check{\mathcal{G}}^{_U}_{_C}\left({\bf k},\omega\right)=-\frac{k_{_0}}{\epsilon_{_0}c^{2}}\;\frac{\hat{\bf k}\otimes\hat{\bf k}}{\epsilon_{_U}\left(k,\omega\right)}.
\end{equation}
(AP): Anisotropic polariton
\begin{align}
&\check{\mathcal{G}}^{_U}_{_{AP}}\left({\bf k},\omega\right)=\frac{k_{_0}}{\epsilon_{_0}c^{2}}\nonumber\\
&\frac{\left(\hat{\bf p}+\epsilon_{_H}\left( k,\theta,\omega\right)\;\hat{\bf k}\right)\otimes\left(\hat{\bf p}+\epsilon_{_H}\;\left( k,\theta,\omega\right)\hat{\bf k}\right)}{k^2-\epsilon_{_E}\left( k,\theta,\omega\right)}.
\end{align}
Taking the trace of the imaginary part of the uniaxial Green function, the FD in reciprocal space is then
\begin{align}
  &\mathcal{F}^{_U}\left({\bf k},\omega\right) = \text{Tr}\left[\text{Im}\left\{\check{\mathcal{G}}^{_U}\left({\bf k},\omega\right)\right\}\right]=\frac{k_{_0}}{\epsilon_{_0}c^{2}}\;\text{Tr}\Bigg[\nonumber\\ 
  &\Bigg(\frac{\text{Im}\left\{\epsilon_{_P}\left( k,\omega\right)\right\}}{|k^2-\epsilon_{_P}\left(k,\omega\right)|^{2}}\;\hat{{\bf s}}\otimes\hat{{\bf s}}+
  \frac{\text{Im}\left\{\epsilon_{_E}\left( k,\theta,\omega\right)\right\}}{|k^2-\epsilon_{_E}\left( k,\theta,\omega\right)|^{2}}\;\hat{\bf p}\otimes\hat{\bf p}+\nonumber\\
  &\Bigg(\left(\frac{c\left(\theta\right)|k^{2}-\epsilon_{_P}\left( k,\omega\right)|}{|\epsilon_{_U}\left( k,\theta,\omega\right)|}\right)^{2}\frac{\text{Im}\left\{\epsilon_{_A}\left( k,\omega\right)\right\}}{|k^2-\epsilon_{_E}\left( k,\theta,\omega\right)|^{2}}+\nonumber\\
  &\left(\frac{s\left(\theta\right)|k^{2}-\epsilon_{_A}\left(k,\omega\right)|}{|\epsilon_{_U}\left( k,\theta,\omega\right)|}\right)^{2}\frac{\text{Im}\left\{\epsilon_{_P}\left( k,\omega\right)\right\}}{|k^2-\epsilon_{_E}\left( k,\theta,\omega\right)|^{2}}\Bigg)\;\hat{\bf k}\otimes\hat{\bf k}\Bigg)\Bigg].
  \label{GreenUniIm}
\end{align}
(The trace of the coupling matrices $\hat{\bf p}\otimes\hat{\bf k}$ and $\hat{\bf k}\otimes\hat{\bf p}$ are zero.) 
The poles of this function show that the first term again represents ordinary (O) excitations, while the final two terms form the combined contributions of the anisotropic polariton (AP) and purely longitudinal (C) families, which couple due the presence of shared longitudinal fields. 
Regardless of this mixing, the influence of these two types of excitations are distinguished by their poles. 
The transverse $\;\hat{\bf p}\otimes\hat{\bf p}$ term has $|k^2-\epsilon_{_E}\left( k,\omega\right)|^{2}$ (AP) type poles, but not $|\epsilon_{_U}\left( k,\omega\right)|^2$ (C) type poles. 
\vspace{-10 pt}
\section{Model-Independent Characteristic}
\vspace{-10 pt}
\noindent
To this point, all permittivity factors have been written as functions of the magnitudes $k$ and $\omega$.
From the inverse Fourier transformation, these dependencies correspond to spatial and temporal nonlocality. 
For most problems in macroscopic electromagnetic, the spatial dependence is dropped, producing 
\begin{align}
&{\bf r}'-{\bf r}\neq {\bf 0}\;\Rightarrow\;  \epsilon\left({\bf r}-{\bf r}', t-t'\right)  = 0, \nonumber \\
& \epsilon\left(k,\omega\right)\rightarrow \epsilon\left(\omega\right). 
\end{align}
For the FD, this approximation raises difficulties.
Ignoring nonlocality removes all spatial dependence of the Coulombic solutions, making the associated longitudinal field scale invariant.
There is then no suppression of arbitrarily high momentum contributions, and the Coulombic piece of the FD diverges. 
(This issue is of course not unique to macroscopic electromagnetics, and is commonly dealt with in quantum field theories by the introduction of regulators, parameter renormalization and counter terms\cite{schwartz2014quantum}.)\\ \\
Yet, while problematic for properties like the excitation lifetime of embedded emitters\cite{Barnett1999}, for isotropic media the local approximation is nevertheless largely valid; and many relevant geometry independent characteristics can be determined from the homogeneous Green function without any treatment of nonlocality. 
This follows from a careful consideration of the length scales involved.
For optical and infrared energies, the wavelength of light is almost always large compared to the scale of nonlocality (e.g. the scale of the material lattice), and the difference between $\epsilon\left( k,\omega\right)$ and $\epsilon\left(\omega\right)$ is usually insubstantial for any externally excited electric field. 
(A fact that also makes accurate probing of the permittivity response at optical and infrared frequencies above $k/k_{_0}\approx 5$\cite{shekhar2017momentum} difficult.)
This approximate indistinguishability of $\epsilon\left( k,\omega\right)$ and $\epsilon\left(\omega\right)$ is equivalent to the assumption that all Coulombic poles can be moved to arbitrarily large reciprocal vector without tangibly altering the Green function for all pertinent reciprocal vectors.
Such a change has has two primary effects.
First, the propagation length of Coulombic solutions becomes vanishingly small.
Second, interaction of an external field with Coulombic solutions becomes possible only in the presence of a physical discontinuity.
Combining these two features, longitudinal fields can only exist at the interface of two media (surface charge densities), and therefore, only influence surface effects.
Volume characteristics, related to the propagation of an external field once it enters the medium\cite{cohen1997photons}, are conversely contained only in the electromagnetic modes, and hence require no knowledge of the Coulombic component.  \\ \\
The mixed fields appearing in AP type modes would seem to preclude a similar separation of domains in anisotropic media.
The asymptotic behavior of \eqref{GreenUniIm} shows that the final term of   
\begin{align}
&\mathcal{F}^{_U}\left(\omega\right)=\nonumber \\
&\lim_{|{\bf r}|\to 0}\int\limits_{0}^{2\pi} d\phi \int\limits_{0}^{\pi} d\theta \int\limits_{0}^{\infty} dk \; \frac{k^{2} s\left(\theta\right)}{\left(2\pi\right)^{3}}\frac{e^{i{\bf k}\cdot{\bf r}}+e^{-i{\bf k}\cdot{\bf r}}}{2}\mathcal{F}^{_U}\left({\bf k},\omega\right)
\label{elecDen}
\end{align}
diverges as $k^{3}$ in the limit of local permittivity response.
Since this term contains contributions from the AP type modes, it is implausible that a meaningful characterization of hyperbolic response can be captured without it.
In turn, this would mean that the FD of a hyperbolic media (or anisotropic media generally) is only describable once a microscopic model (or other physically motivated normalization) is specified. \\ \\
\noindent
To work around this apparent difficulty, we begin by expanding all absolute values and imaginary parts of \eqref{GreenUniIm}, treating $k$ as a real variable, and extending the resulting expression over the entire complex $k$ plane.
Jordan's lemma and the Cauchy integral theorem then imply that if $k^{2}\mathcal{F}^{_U}\left({\bf k},\omega\right)$ tends to a constant value as $|k|\to \infty$ the value of $\mathcal{F}^{_U}\left(\omega\right)$ equals the residues of $\mathcal{F}^{_U}\left({\bf k},\omega\right)$.
(Here, we are taking infinite semi-circle contours in the upper and lower half spaces depending on the value of ${\bf k}\cdot{\bf r}$, with the convention that the square root function is cut along the positive real axis.)
Assuming nonlocality leads to $\text{Im}\left\{\epsilon_{_A}\left(k,\omega\right)\right\}$ and $\text{Im}\left\{\epsilon_{_P}\left(k,\omega\right)\right\}$ having scaling at least $\propto 1/k^{2}$, so that there is no contribution from the path at arbitrarily large $k$, these residues split into two distinct classes.\\ \\
(i) Poles of the form $k_{_S}=\pm\sqrt{\epsilon_{_X}\left(k_{_S},\omega\right)}$ resulting from the wave equations that have a second order dependence on spatial nonlocality, i.e. poles that tend to $k_{_S}=\pm\sqrt{\epsilon_{_X}\left(\omega\right)}$ in the limit of local response. (Here $k_{_S}$ and $\epsilon_{_X}$ are place holder labels that could apply to either ordinary or AP modes.) \\ \\
(ii) Poles that have a first order dependence on the exact characteristics of nonlocality, and thus tend to towards arbitrarily large values in the limit of local response.\\ \\
Following the same reasoning as isotropic media, volume characteristics, for low to moderate $k$ fields, must depend only on the first class pole. 
Therefore, we can conclude that the sum of these residues is the correct generalization of the transverse FD of an isotropic material to an anisotropic setting.\\ \\  
The validity of the above argument rests on the asymptotic $k$ scaling of the imaginary part of permittivity response being stronger than $\propto 1/k^{2}$, but this is a generally valid assumption for any medium.
As $\epsilon\left( k,\omega\right)-1=\chi\left( k,\omega\right)$ is a response function, it is analytic for all but an finite set of points in the complex $k$ plane for a given $\omega$\cite{abrikosov1975ie}. 
Correspondingly, there is a convergent Laurent series expansion in complex $k$ such that 
\begin{align}
  \chi\left( k,\omega\right)=\sum\limits_{n=-\infty}^{\infty}c_{n}\left(\omega\right) k^{n}
\end{align}
for $M < |k| <\infty$, where $M$ is magnitude of the largest $\bf k$ pole of $\chi\left( k,\omega\right)$, and $\left\{c_{n}\left(\omega\right)\right\}$ are the frequency dependent coefficients of the expansion. 
From the definition of the Fourier transform, $\chi\left( k,\omega\right)$ is related to the real space susceptibility as 
\begin{align}
  \chi\left( r,\omega\right)=\int\limits_{-\infty}^{\infty} dk\; e^{i k r} \chi\left(k,\omega\right),
\end{align} 
and the required convergence of this expression for all values of $r$ is only guaranteed if $\left(n\geq 1\right)\Rightarrow c_{n}\left(\omega\right)=0$. 
(Although we have assumed inversion symmetry, $\epsilon\left({\bf k},\omega\right)=\epsilon\left(k,\omega\right)$, throughout the manuscript all arguments and forms we present hold generally.) \\ \\
With this result in hand, the residues determined by the wave equation poles sum to give 
\begin{align}
  &\mathcal{F}^{_U}_{_O}\left(\omega\right)=\int\limits_{0}^{2\pi}d\phi\int\limits_{0}^{\pi/2}d\theta\;s\left(\theta\right) \text{Tr}\left[\text{Im}\left\{\check{\mathcal{G}}^{_U}_{_{O}}\left(\theta,\phi,\omega\right)\right\}\right]=\nonumber \\
  &\frac{k_{_0}\pi}{\left(2\pi\right)^{3}\epsilon_{_0}c^{2}}\int\limits_{0}^{2\pi}d\phi\int\limits_{0}^{\pi/2}d\theta\;s\left(\theta\right)\;Re\left\{\sqrt{\epsilon_{_P}\left( k_{_O},\omega\right)}\right\}
  \label{GreenO}
\end{align}
for O type modes, and 
\begin{align}
  &\mathcal{F}^{_U}_{_{AP}}\left(\omega\right)=\int\limits_{0}^{2\pi}d\phi\int\limits_{0}^{\pi/2}d\theta\;s\left(\theta\right) \text{Im}\left\{\check{\mathcal{G}}^{_U}_{_{AP}}\left(\theta,\phi,\omega\right)\right\}=\nonumber\\
  &\frac{k_{_0}\pi}{\left(2\pi\right)^{3}\epsilon_{_0}c^{2}}\int\limits_{0}^{2\pi}d\phi\int\limits_{0}^{\pi/2}d\theta\;s\left(\theta\right)\;Re\left\{\sqrt{\epsilon_{_E}\left( k_{_E}\left(\theta\right),\theta,\omega\right)}\right\}\nonumber\\
  &\Bigg(1+\left|\epsilon_{_H}\left(k_{_E}\left(\theta\right),\theta,\omega\right)\right|^{2}\Bigg)
  \label{GreenE}
\end{align}
for AP type modes, with $k_{_O}$ and $k_{_E}\left(\theta\right)$ standing for the solutions to \eqref{ordCond} and \eqref{eoCond} nearest those of the local approximation in the complex plane. In \eqref{GreenE}, the first term results from the transverse field, and the second from the longitudinal field. \\ \\
Convincingly, \eqref{GreenO} and \eqref{GreenE} are also precisely the result obtained by considering the normal variables found in the second section. 
Directly, \eqref{GreenUni} identifies the transverse part of the AP excitations with the second term of \eqref{GreenUniIm}. 
Since there is no question as to the convergence of this term, the existence of \eqref{GreenE} (the longitudinal part) must follow as a requirement of the Maxwell equations.
(Either approach to \eqref{GreenE} is independent of the specific form of nonlocality considered.) 
Further, the permittivity dependence of the Coulombic piece of this term  
\begin{align}
  &\mathcal{F}^{_U}_{_{AP}\;_{L}}\left(\omega\right)=
  \frac{k_{_0}\pi}{\left(2\pi\right)^{3}\epsilon_{_0}c^{2}}\int\limits_{0}^{2\pi}d\phi\int\limits_{0}^{\pi/2}d\theta\;s\left(\theta\right)\nonumber\\
  &Re\left\{\sqrt{\epsilon_{_E}\left( k_{_E}\left(\theta\right),\theta,\omega\right)}\right\}\left|\epsilon_{_H}\left(k_{_E}\left(\theta\right),\theta,\omega\right)\right|^{2}
\end{align}
is the same as that observed in calculating the power radiated by a dipole in a losses hyperbolic medium\cite{Felsen1994,Zhukovsky2013}, and is essentially an angular version of the FD associated with a surface plasmon polariton excitation\cite{ford1984electromagnetic}.
(Note that in this contribution the usual distinctions between one and two-sheeted hyperboloids\cite{Cortes2012} - metallic response along the optical axis, or in the optical plane - are nearly absent.)
\\ \\
\noindent
Computation of the FD resulting from the second class of poles, on the other hand, does requires a specific model of nonlocality\cite{Horsley2014, pollard2009optical} (just as in isotropic media). 
To focus our discussion we will not pursue these details. 
Still, there are some general characteristics worth noting. 
Considering the longitudinal component of \eqref{GreenUniIm}, once $k^2$ surpasses $| Re\left\{\epsilon_{_E}\left(k,\theta,\omega\right\}\right) |$ the prefactors $\left(|k^{2}-\epsilon_{_P}\left( k,\omega\right)|/|k^2-\epsilon_{_E}\left( k,\theta,\omega\right)|\right)^{2}$ and $\left(|k^{2}-\epsilon_{_A}\left( k,\omega\right)|/|k^2-\epsilon_{_E}\left( k,\theta,\omega\right)|\right)^{2}$ will both quickly approach unity. 
(Assuming nonlocality does not drastically increase the peak magnitude of the polarization response for real $k$.) 
Once this condition is achieved, the final term of \eqref{GreenUniIm} is increasingly well approximated as 
\begin{equation}
  \begin{split}
  \text{Im}\left\{\check{\mathcal{G}}^{_U}_{_C}\left({\bf k},\omega\right)\right\}=\frac{k_{_0}}{\epsilon_{_0}c^{2}}\Bigg(
  \frac{\hat{\bf k}\otimes\hat{\bf k}}{|\epsilon_{_U}\left( k,\theta,\omega\right)|^2} \text{Im}\left\{\epsilon_{_U}\left( k,\theta,\omega\right)\right\}\Bigg).
  \label{GreenCou}
  \end{split}
\end{equation}
This is again the exact result found by considering the normal variables of the Coulombic solutions independently, and a straightforward extension of the Coulombic FD encountered in isotropic media,
\begin{equation}
  \begin{split}
  \text{Im}\left\{\check{\mathcal{G}}^{_I}_{_C}\left({\bf k},\omega\right)\right\}=\frac{k_{_0}}{\epsilon_{_0}c^{2}}\Bigg(
  \frac{\hat{\bf k}\otimes\hat{\bf k}}{|\epsilon\left( k,\omega\right)|^2} \text{Im}\left\{\epsilon\left( k,\omega\right)\right\}\Bigg).
  \label{GreenCouI}
  \end{split}
\end{equation}
For $k$ where the above approximation is valid, the residues from these (C) type poles can safely be attributed to pure Coulombic modes.\\ \\
Running contrary to this discussion, it should be emphasized that in cases where strong interactions with Coulombic solutions are expected (for example embedded emitter) that the Green function forms provided previously, in particular \eqref{GreenUniIm}, are well suited to computation. 
They are valid in any general uniaxial (or isotropic) medium, and can be normalized using any specific model of non-locality or standard quantum field theory approach.  
\vspace{-10 pt}
\section{The Sum Rule for Modified Spontaneous Emission Enhancement in Hyperbolic Media}
\vspace{-10 pt}
\begin{figure*}[ht!]
  \centering
   \vspace{0 pt}
  \includegraphics{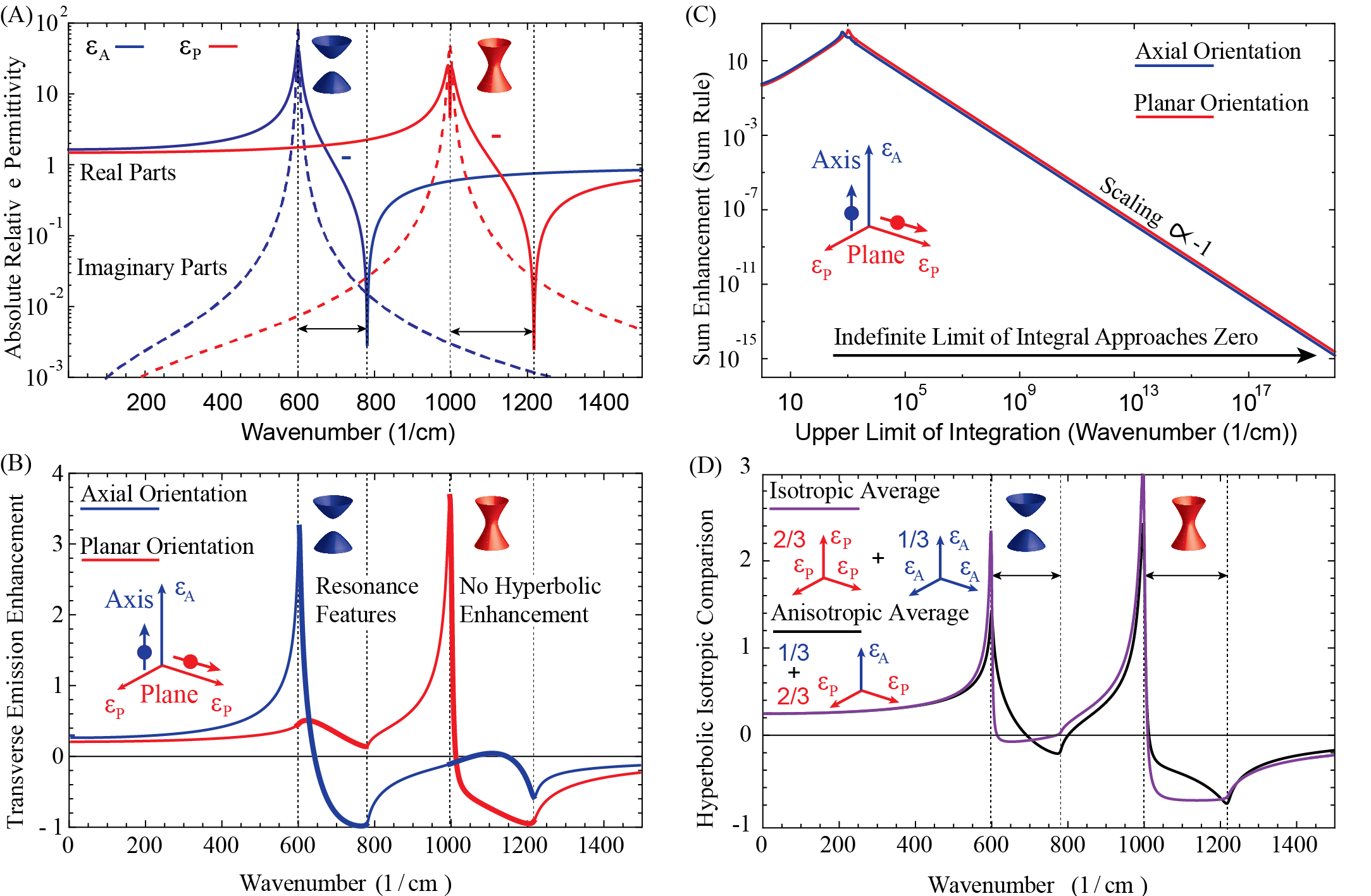}
  \vspace{-10 pt}
  \caption{\textbf{Sum Rule for Transverse Spontaneous Emission Enhancement in Hyperbolic Media}\\  Panel (A) displays the absolute relative permittivity values resulting from \eqref{polModel}. The thin dashed lines and schematic dispersion surfaces highlight spectral regions of hyperbolic response where one of either $Re\left\{\epsilon_{_P}\left(\omega\right)\right\}$ or  $Re\left\{\epsilon_{_A}\left(\omega\right)\right\}$ is negative. Panel (B) shows the resulting transverse spontaneous emission enhancement offset by vacuum, the integrand of \eqref{sumRule}. Panel (C) plots the integrated enhancement as function of the upper wavenumber considered. These result confirm that the enhancement sum rule is strictly obeyed inside hyperbolic media\cite{Scheel2008}. Accounting only for purely electromagnetic (transverse) contributions, emission enhancement in spectral regions of hyperbolic response is unremarkable. Panel (D) further highlights this fact by comparing the orientationally averaged enhancement from (B), black line, with the enhancement found by averaging two isotropic media with $\epsilon\left( k,\omega\right)=\epsilon_{_P}\left(\omega\right)$ and $\epsilon\left( k,\omega\right)=\epsilon_{_A}\left(\omega\right)$ weighted by factors of $2/3$ and $1/3$ respectively. The graphs are found to be nearly identical, even though the two situations correspond to very different electromagnetic environments.}
  \vspace{-10 pt}
\end{figure*}
\noindent
The sum rule for modified spontaneous emission enhancement, formulated by Barnett and Loudon\cite{Barnett1996,barnett1998sum}, states that it is not possible to alter the total relative rate of spontaneous emission into purely electromagnetic (transverse) modes.
If the properties of a medium enhance the relative rate of spontaneous emission in one spectral range, they must equally suppress this relative rate in another. Mathematically, this is written as
\begin{equation}
  \int_{0}^{\infty}d\omega\;\frac{\Gamma_{_T}\left({\bf r},\omega\right)-\Gamma_{_0}\left(\omega\right)}{\Gamma_{_0}\left(\omega\right)}=0,
  \label{sumRule}
\end{equation}
where
\begin{equation}
  \Gamma_{_T}\left({\bf r},\omega\right)=\frac{2\omega^{2}}{\hbar}{\bf d}\;\text{Im}\left\{\pi_{_T}\check{G}\left({\bf r},{\bf r},\omega\right)\pi_{_T}\right\}\;{\bf d}
  \label{relativeEmission}
\end{equation}
is the relative rate of spontaneous emission of a single level emitter of frequency $\omega$, with transition dipole moment ${\bf d}$, at position ${\bf r}$ in a medium described by $\check{G}\left({\bf r},{\bf r}',\omega\right)$, $\pi_{_T}$ is the transverse projection operator, and 
\begin{equation}
  \Gamma_{_0}\left({\bf r},\omega\right)=\frac{k_{_0}^{3}}{3\pi\hbar\epsilon_{_0}}\;{\bf d}\;\check{I}\;{\bf d}
  \label{gammao}
\end{equation}
is the rate of spontaneous emission in vacuum.
\begin{figure}[ht!]
  \centering
   \vspace{0 pt}
  \includegraphics{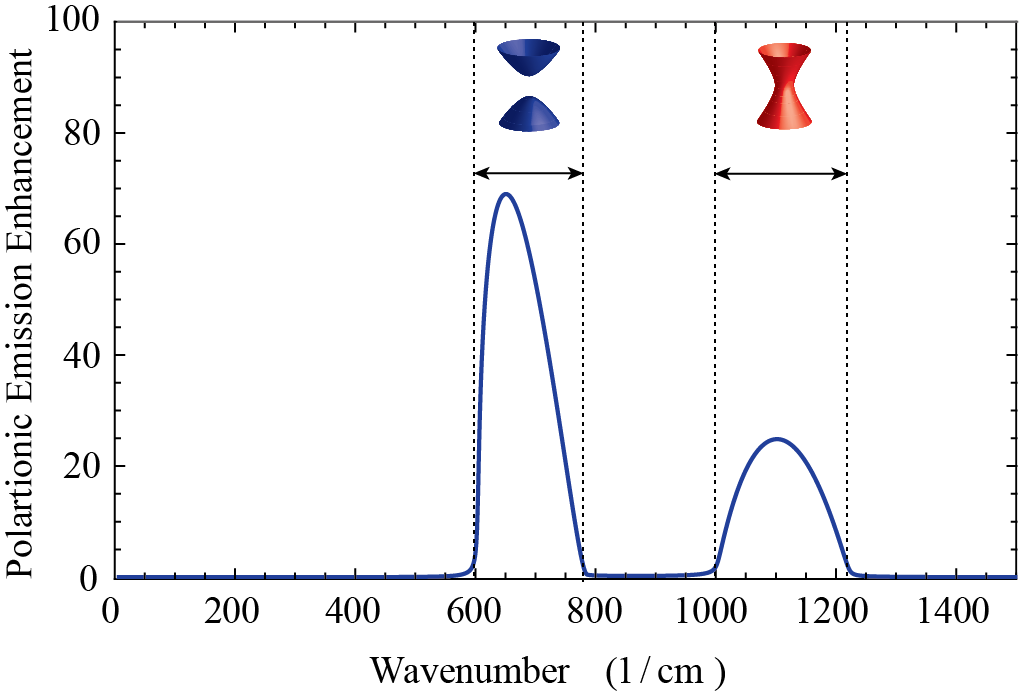}
  \vspace{-10 pt}
  \caption{\textbf{Polaritonic Spontaneous Emission Enhancement Surpassing the Sum Rule in Hyperbolic Media}\\ The figure displays the orientationally averaged spontaneous emission enhancement of the longitudinal part of the AP contribution, considering a uniaxial media with permittivity model \eqref{polModel}. Contrasting with Fig.2, this longitudinal enhancement dominants in spectral regions of hyperbolic response.}
  \vspace{0 pt}
\end{figure}
\vspace{0 pt}
\\ \\
\noindent
As anisotropy does not change the frequency dependence of the Green function, an argument for the general validity of the sum rule, given by Scheel\cite{Scheel2008}, theoretically guarantees that the transverse part of the uniaxial Green function 
\begin{align}
  &\text{Im}\left\{\pi_{_T}\check{G}^{_U}\left(\omega\right)\pi_{_T}\right\}=\frac{k_{_0}\pi}{\left(2\pi\right)^{3}\epsilon_{_0}c^{2}}\int\limits_{0}^{2\pi}d\phi\int\limits_{0}^{\pi/2}d\theta\;s\left(\theta\right) \nonumber \\
  &\left(Re\left\{\sqrt{\epsilon_{_P}\left(\omega\right)}\right\}\;\hat{\bf s}\otimes\hat{\bf s}+Re\left\{\sqrt{\epsilon_{_E}\left(\theta,\omega\right)}\right\}\;\hat{\bf p}\otimes\hat{\bf p}\right),
  \label{GreenTrans}
\end{align}
must satisfy \eqref{sumRule} so long as the permittivities considered satisfy the Kramers-Kronig relations. An illustrative example of this result, assuming local Lorentzian polarization responses
\noindent
\begin{equation}
  \epsilon\left(\omega\right)=1+\frac{\omega_{\rho}^{2}}{\omega_{_0}^2-\omega\left(\omega+i\gamma\right)},
  \label{polModel}
\end{equation}
for $\epsilon_{_{A}}\left(\omega\right)$ and $\epsilon_{_{P}}\left(\omega\right)$ with $\omega_{\rho}=\left\{ 500_{_A},\; 700_{_P}\right\}\; cm^{-1}$, $\omega_{_0}=\left\{ 600_{_A}\;, 1000_{_P}\right\}\; cm^{-1}$ and $\gamma=\left\{ 5_{_A}\;, 10_{_P}\right\}\; cm^{-1}$ is provided in Fig.2. 
From the graph, we observe that the regions of hyperbolic response are essentially featureless, and, as in isotropic media, enhancement follows the magnitude of polarization.
In fact, Fig.2(D) shows that the orientationally averaged enhancement of this transverse part is nearly equivalent to considering the planar and axial permittivities separately and summing the result. 
(Precisely, replacing \eqref{GreenTrans} with the sum of $(2/3)\;\mathcal{F}^{I}_{_{T}}\left(\omega\right)$ with $\epsilon\left( k,\omega\right)=\epsilon_{_P}\left(\omega\right)$ and $(1/3)\; \mathcal{F}^{I}_{_{T}}\left(\omega\right)$ with $\epsilon\left( k,\omega\right)=\epsilon_{_A}\left(\omega\right)$, produces a result that is not significantly different form that of the true hyperbolic medium.)
\noindent \\ \\
These results for the transverse enhancement of the spontaneous emission, which hold to arbitrarily low absorption ($\gamma$), follow from the normal variable picture.
Any property of a linear macroscopic medium should be consistent with some arrangement of dipoles in a vacuum. 
Since scattering from a dipole does not introduce new electromagnetic modes, there is no way that the sum can be modified. 
From \eqref{GreenE} the resonant effects of hyperbolic response for AP type excitations occur in the Coulombic field\cite{cortes2017super}. 
In taking the strictly electromagnetic (transverse) part of \eqref{GreenE} these features are ignored.\\ \\
\noindent
Orientationally averaged spontaneous emission enhancement resulting from the Coulombic portion of the AP FD
\begin{align}
  &\frac{\Gamma^{_U}_{_{AP\;L}}\left(\omega\right)}{\Gamma_{_0}\left(\omega\right)}=\frac{6\pi\epsilon_{_0}c^{3}}{\omega} \mathcal{F}^{_U}_{_{AP\;L}}\left(\omega\right)=\nonumber\\
  &\frac{3}{2}\int\limits_{0}^{\pi/2}d\theta\;s\left(\theta\right)
  Re\left\{\sqrt{\epsilon_{_E}\left(\theta,\omega\right)}\right\}\left|\epsilon_{_H}\left(\theta,\omega\right)\right|^{2},
\label{GreenLong}
\end{align}
is plotted in Fig.3. (For numerical convenience in the remainder of this article the FD will be taken to be vacuum normalized by the prefactor appearing in \eqref{GreenLong}.) 
Comparing with Fig.2, it is clear that this enhancement does not obey the sum rule: it is an additional positive contribution that grows arbitrarily large as material absorption is decreased.
By itself, this fact is not particularly unusual. 
In a general isotropic medium the absorption of energy into matter is not limited by the number of electromagnetic modes \eqref{sumRule}, and so neither is the enhancement contribution of Coulombic modes. 
Yet, there are key distinctions between these two cases.
\\ \\
(i) The AP enhancement of the FD does not diverge in the limit of local permittivity response (nonlocality is a second order effect). 
This is not the case for purely Coulombic enhancement\cite{Horsley2014}.\\ \\
(ii) The AP enhancement of the FD is not simply related to the magnitude of the polarization, or total density of charge carriers, as has been shown for Coulombic enhancement in isotropic media\cite{Scheel2008}.
Instead, it depends principally on the magnitude of anisotropy, $|\epsilon_{\Delta}\left(\omega,\theta\right)|^{2}$ and material absorption, $\left(Re\left\{\epsilon_{_U}\left(\omega,\theta\right)\right\} |\epsilon_{_U}\left(\omega,\theta\right)|^{2}\right)^{-1}$.\\ \\  
(iii) The AP type enhancement of the FD shows a unique scaling with material absorption, which is stronger than the scaling exhibited by either the transverse O type \eqref{GreenTrans} or longitudinal C type \eqref{GreenCou} enhancement, Fig.4.\\ \\
These final two points are salient for potential applications involving embedded quantum emitters.
Although nothing can be stated unequivocally without knowledge of the nonlocal response features, the scaling trends indicate that polaritonic solutions become dominate if either the material loss, $Im\left\{\epsilon\right\}$, is small, or the anisotropy is large. 
\begin{figure}[ht!]
  \centering
   \vspace{-5 pt}
  \includegraphics{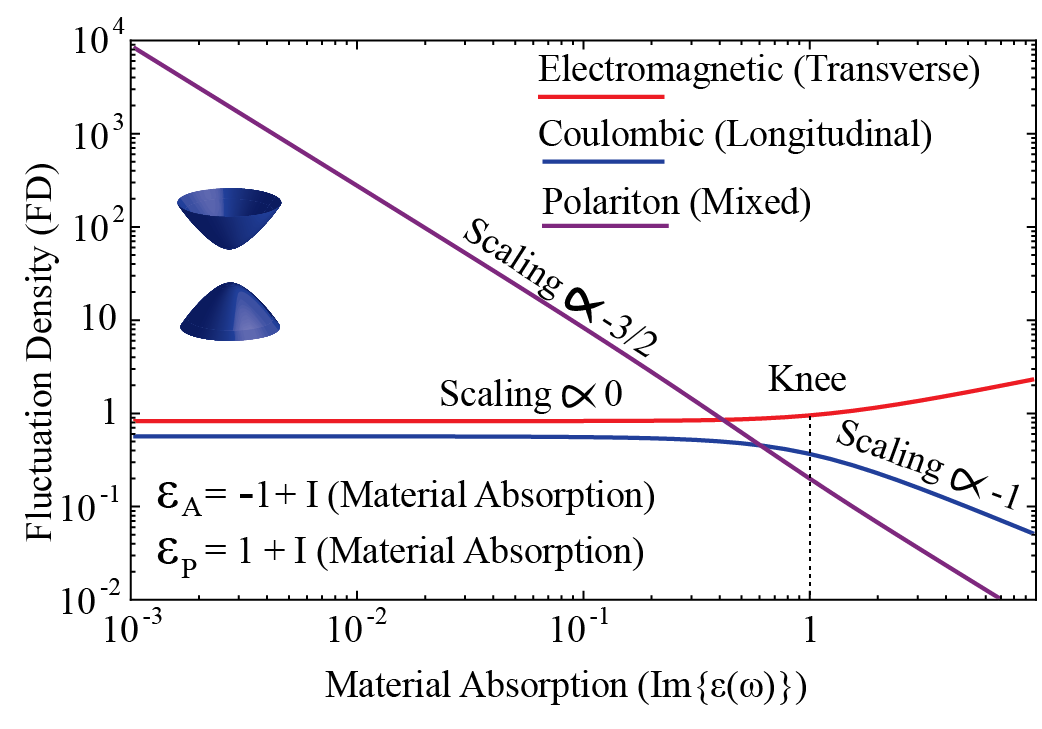}
  \vspace{-15 pt}
  \caption{\textbf{Scaling with Material Absorption of Fluctuation Densities in Hyperbolic Media }\\ The figure depicts the power scaling of the transverse O type \eqref{GreenTrans}, longitudinal C type \eqref{GreenCou}, and mixed AP type (longitudinal part only) \eqref{GreenLong} contributions to the FD as a function of material absorption $\text{Im}\left\{\epsilon_{_A,_P}\left(\omega\right)\right\}$. For the C type contribution, only the angular integrals in \eqref{elecDen} have been computed as the $k$ integral diverges in the limit of local polarization response. The plot is computed by considering $\epsilon_{_A}=-1 + i\left(\text{x-axis value}\right)$, $\epsilon_{_P}=1 + i\left(\text{x-axis value}\right)$. The knee transitioning from a scaling of $\propto -1$ to a scaling of $\propto 0$ is set by the minimum magnitude of the real permittivity components $|Re\left\{\epsilon_{_A,_P}\left(\omega\right)\right\}|$. (This behaviour is also seen the isotropic case\cite{Barnett1999}.) The $x^{-3/2}$ scaling exhibited by the AP contribution is found to be stronger than either of the two pure solution types.}
  \vspace{-10 pt}
\end{figure}
\vspace{-10 pt}
\section{Thermal Fluctuations in Hexagonal Boron Nitride and Bismuth Selenide}
\vspace{-10 pt}
\noindent
Like the degree of relative spontaneous emission enhancement, the thermal energy density in the electric and magnetic fields is similarly set by the FD through the relation
\begin{align}
    &\mathcal{U}\left({\bf r},\omega,T\right)=\frac{\epsilon_{_0}}{2}\text{Tr}\left[\left<{\bf E}\left({\bf r},\omega\right)\otimes {\bf E}^{*}\left({\bf r},\omega\right)\right>\right]+\nonumber\\
    &\frac{1}{2\mu_{_0}}\text{Tr}\left[\left<{\bf B}\left({\bf r},\omega\right)\otimes {\bf B}^{*}\left({\bf r},\omega\right)\right>\right],
\label{energyTot}
\end{align}
\begin{figure*}[ht!]
  \centering
  \includegraphics{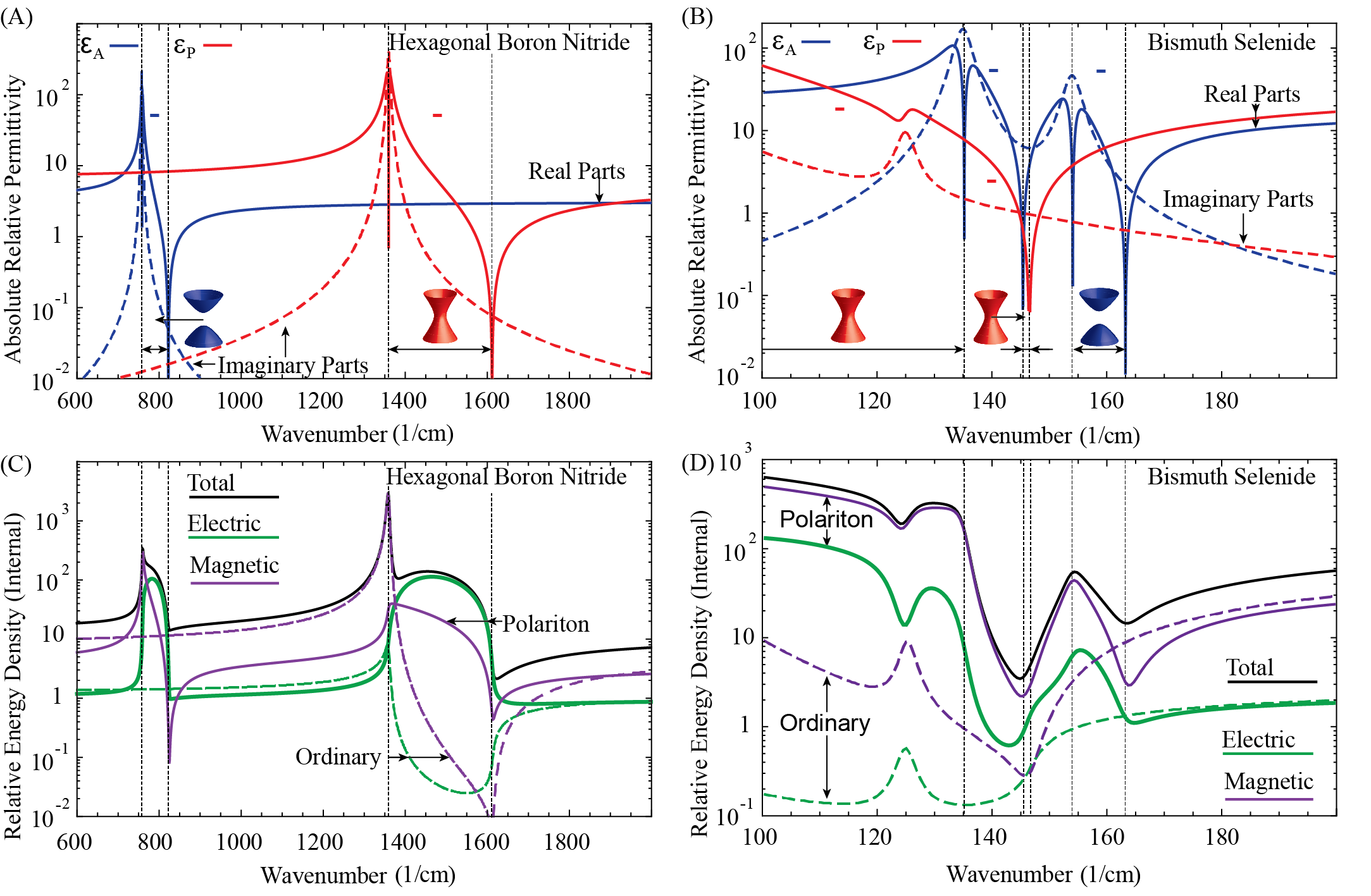}
  \vspace{-20 pt}
  \caption{\textbf{Relative Thermal Energy and Fluctuation Densities in Natural Hyperbolic Media}\\ The figure shows the contribution that AP type modes, solid lines, and O type modes, dashed lines, make to the total electric and magnetic thermal energy densities (total lines) inside hexagonal boron nitride (C) and bismuth selenide (D). (For comparison the energy densities are normalized by half the thermal energy density of vacuum.) The absolute value of the real and imaginary parts of relative permittivity components of these two materials, based on data from references\cite{Caldwell2014,Wu2015}, are plotted in figures (A) and (B). Each sharp peak and dip in (B) signals a sign flip of the real part as marked. The imaginary part of each component remains positive throughout. The green electric lines (bold polariton, dashed ordinary) double as the respective FDs. Both media show broad spectral regions where this quantity is over 120 times larger than it is in vacuum.}
  \vspace{0 pt}
\end{figure*}
\noindent
with 
\begin{equation}
  \text{Tr}\left[\left<{\bf E}\left({\bf r},\omega\right)\otimes {\bf E}^{*}\left({\bf r},\omega\right)\right>\right]=\frac{\omega\;\Theta\left(\omega,T\right)}{\pi}\mathcal{F}\left(\omega\right).
  \label{traceFluc}
\end{equation}
\noindent
\begin{figure}[ht!]
\vspace{0 pt}
  \centering
  \includegraphics{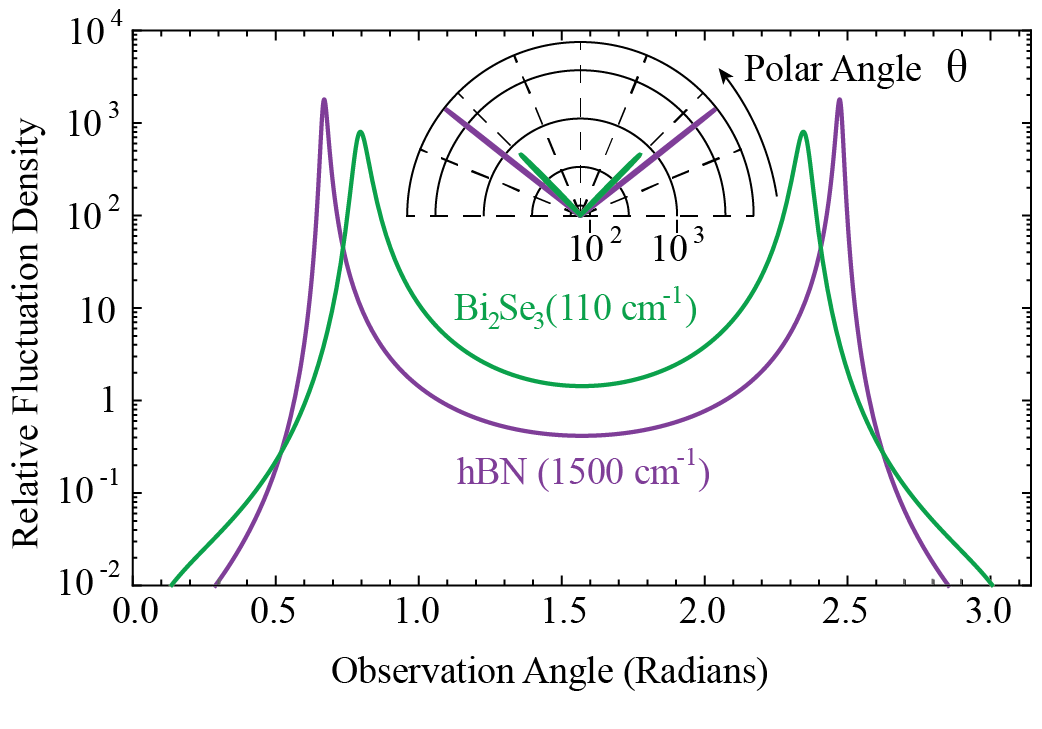}
  \vspace{-20 pt}
  \caption{\textbf{Angular Polaritonic Fluctuation Density in Natural Hyperbolic Media}\\ The figure depicts the polaritonic FD inside hexagonal boron nitride at 1500 $cm^{-1}$, and bismuth selenide at 110 $cm^{-1}$, as a function of polar angle on a logarithmic scale. The inset shows this same quantity as a polar plot on a linear scale. Although the integrated FDs of these two cases are almost equal, Fig.6, as hexagonal boron nitride more closely approaches the resonance condition $|\epsilon_{_U}\left(\theta,\omega\right)|=0$, but possess less polarization anisotropy, the angular distribution of its FD is much more radical. Both materials show angular regions where the relative polar FD is over 800 times larger than vacuum.}
  \vspace{-10 pt}
\end{figure}\\ \noindent
Using \eqref{flucDiss}, \eqref{GreenO}, \eqref{GreenE}, and (6c), the electromagnetic energy density in the O and AP type modes of a uniaxial medium is then
\begin{equation}
  \mathcal{U}\left({\bf r},\omega,T\right)=\frac{\mathcal{U}_{BB}\left(\omega,T\right)}{2}\left(\mathcal{F}_{_E}\left(\omega\right)+\mathcal{F}_{_M}\left(\omega\right)\right),
\end{equation}
with 
\begin{align}
    &\mathcal{F}_{_E}\left(\omega\right) = Re\left\{\sqrt{\epsilon_{_P}\left(\omega\right)}\right\}+\nonumber\\
    &\int\limits_{0}^{\pi/2}d\theta\;s\left(\theta\right)Re\left\{\sqrt{\epsilon_{_E}\left(\theta,\omega\right)}\right\}\left(\left|\epsilon_{_H}\left(\theta,\omega\right)\right|^{2}+1\right)
    \label{elecEnergy}
\end{align}
and 
\begin{align}
    &\mathcal{F}_{_M}\left(\omega\right) = \left|\epsilon_{_P}\left(\omega\right)\right|
    Re\left\{\sqrt{\epsilon_{_P}\left(\omega\right)}\right\}+\nonumber\\
    &\int\limits_{0}^{\pi/2}d\theta\;s\left(\theta\right)\left|\epsilon_{_E}\left(\theta,\omega\right)\right|Re\left\{\sqrt{\epsilon_{_E}\left(\theta,\omega\right)}\right\}
    \label{magEnergy}
\end{align}
denoting the relative electric and magnetic contributions, and $\mathcal{U}_{BB}\left(\omega,T\right)$ the energy density of an ideal blackbody.
\\ \\
\noindent
The results of this expression (ignoring pure Coulombic contributions) for hexagonal boron nitride and bismuth selenide, normalized by $\mathcal{U}_{BB}\left(\omega,T\right)/2$ for direct comparison with the FD, are plotted in Fig.5. Four aspects of this figure warrant attention.  
\vspace{0 pt}
\begin{figure*}[ht!]
  \centering
  \includegraphics{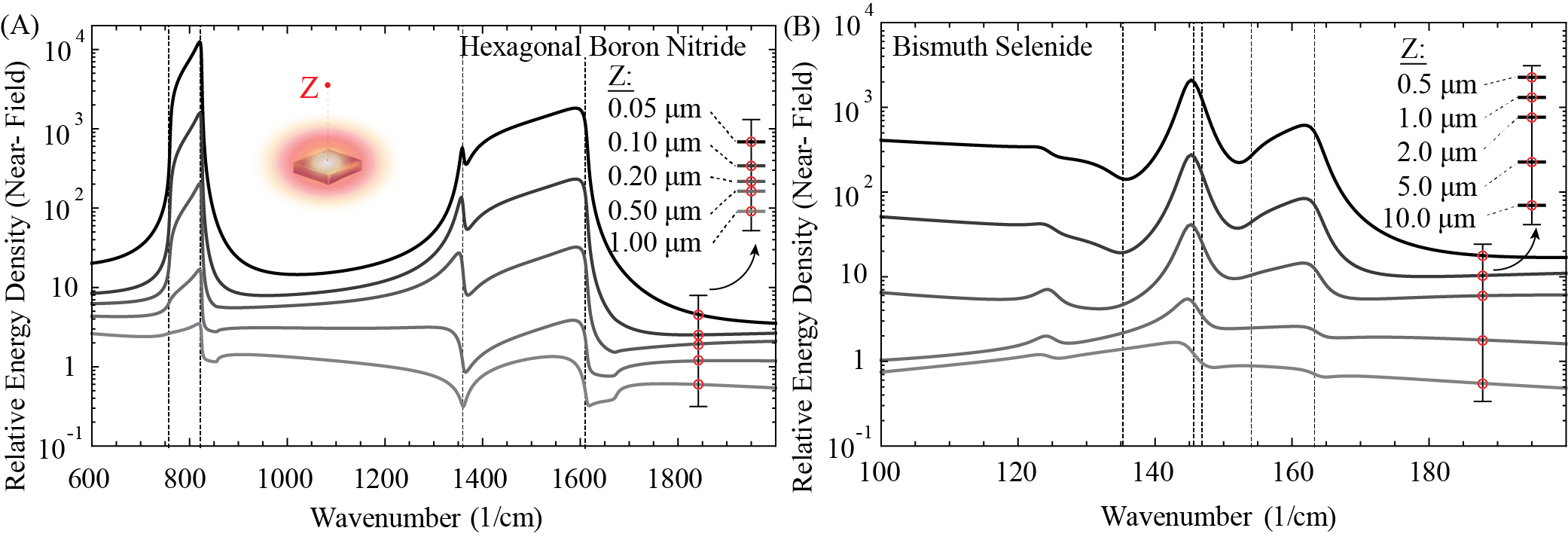}
  \vspace{0 pt}
  \caption{\textbf{Near-Field Electromagnetic Energy Density of Natural Hyperbolic Media}\\ The figure plots the sum of the near-field electric and magnetic energy densities above half-spaces of hexagonal boron nitride (A) and bismuth selenide (B) for increasingly small observation distances. (For comparison with the FD results this energy is normalized by half the energy density of vacuum.) The inset in panel (A) shows a schematic representation. Recalling that surface mode contributions are not included in \eqref{energyTot}, the relative spectral characteristics of the near-field energy density are nevertheless seen to be in good agreement with FDs plotted in Fig. 5 for small observation distances.}
  \vspace{0 pt}
\end{figure*}\\ \\
\noindent
(i) The solid green curves, denoting the contribution of AP type modes, confirm that in real media either a high degree of anisotropy $\epsilon_{_\Delta}\left(\theta,\omega\right)$ or low material absorption $\epsilon_{_U}\left(\theta,\omega\right)$ may lead to a large polaritonic FD.
Bisumth selenide exhibits substantial material absorption, yet nevertheless, large enhancement results from the extreme difference between the axial and planar permittivity components. 
Hexagonal boron nitride possess much less anisotropy, but has lower material absorption, leading to a nearly identical FD.\\ \\
\noindent
(ii) The energy density of the magnetic field is often roughly an order of magnitude larger than that of the electric field (total lines).\\ \\
\noindent 
(iii) \eqref{traceFluc} equates the green lines with the FD contributions of AP (bold line) and O (dashed line) type modes. 
Both hexagonal boron nitride and bismuth selenide show broad spectral regions where the FD is over 120 times larger than vacuum.\\ \\
\noindent
(iv) Moving to Fig.6, the AP component of the FD has extreme angular dispersion, concentrated along the critical angles determined by $Re\left\{\epsilon_{_U}\left(\theta,\omega\right)\right\}=0$. 
Along the cone set by this angle, the polaritonic FD is over 800 times larger than vacuum FD in both bismuth selenide and hexagonal boron nitride. \\ \\
Given the degree of enhancement observed, one may question whether the use of local response in \eqref{elecEnergy} and \eqref{magEnergy} is accurate. 
Without experimental evidence, this is open; but we are inclined to believe that the approximation does hold. 
In either material, the largest absolute value of the permittivities $\epsilon_{_A}\left(\omega\right)$, $\epsilon_{_P}\left(\omega\right)$, and $\epsilon_{_E}\left(\omega\right)$ is $\approx 400$, and the smallest $\approx 0.1$. 
Based on these bounds, in the local approximation, all wave equation poles occur below $20 \;k_{o}$. 
Taking the largest lattice spacing present in either material, $\approx 3\; nm$, this upper limit of $k$ still corresponds to less than $1\%$ of the Brillouin zone for wavelengths longer than $6\; \mu m$. 
(The smallest wavelength of hyperbolic response considered occurs in hexagonal boron nitride at $6.25\;\mu m$.) 
As such a small change will only minimally modify the probed bandstructure around the dominant optical phonon features\cite{Serrano2007, Zhang2010, Wu2015}, substantial variation of the permittivity response should not be expected.
\\ \\
It is interesting to compare the electromagnetic energy plotted in Fig.5 with the full near-field energy density above a half space of hyperbolic media\cite{Guo2012,guo2014fluctuational}. 
For this purpose, equation (16) of Guo et al.\cite{guo2014fluctuational} is plotted in Fig.7. 
(Other examples of calculated thermal properties that can be compared with volume FDs appear in a recent work by Liu et al.\cite{liu2016super}.) 
At the nearest observation points considered, the near-field calculation produces values larger than the associated bulk value, signaling that enhancement at these distances is driven primarily by the excitation of surface charges. 
All the same, recalling that in Fig.5 no C type or surface polariton modes are included, the two figures show good agreement. 
The additional peaks seen in Fig.7 match the surface polariton condition $Re\left\{\epsilon\left( k,\omega\right)\right\}=-1$. 
This observation indicates that \eqref{GreenO} and \eqref{GreenE} are the correct measures of local electromagnetic fluctuation characteristics in a hyperbolic medium, akin to the index of an isotropic medium. 
Further support of this claim is seen in experimentally reported confinement factors for hexagonal boron nitride resonators\cite{Caldwell2014}, which are within $30~\%$ of the FD we have found for hexagonal boron nitride at these wavelengths. 
\vspace{0 pt}
\section{Summary}
\vspace{-10 pt}
\noindent
In summary, we have shown that material absorption analytically quantifies electromagnetic fluctuations in hyperbolic (generally uniaxial) media in a manner entirely analogous to the isotropic media.
From this result, we have studied the sum rule for modified spontaneous emission enhancement, and have found that it does not apply to the key polaritonic features of a hyperbolic response. 
We have also investigated the density of electromagnetic fluctuations (electromagnetic thermal energy density) inside both hexagonal boron nitride and bismuth selenide. 
We have found that both media have broad spectral regions where this quantity is over 120 times (along specific angular directions 800 times) larger than it is in vacuum. 
Our results unify the computation of electric field fluctuations in uniaxial and isotropic settings, and should prove useful for testing the potential of hyperbolic systems in emerging optical technologies. 
\vspace{-10 pt}
\section*{Acknowledgments}
\vspace{-10 pt}
\noindent
This work has been supported by funding from the National Science and Engineering Research Council of Canada, Alberta Innovates Technology Futures, and the National Science Foundation.
We thank Cris Cortes and Prashant Shekhar for valuable discussions in preparing the article, as well as Chensheng Gong and Todd Van Mechelen for their contributions to the initial investigation of the ideas we have pursued here.
\bibliography{natHyp2}
\end{document}